# Quantum potential energy as concealed motion

Peter Holland

Green Templeton College
University of Oxford
Oxford OX2 6HG
England

peter.holland@gtc.ox.ac.uk

**Abstract**

It is known that the Schrödinger equation may be derived from a hydrodynamic model in which the Lagrangian position coordinates of a continuum of particles represent the quantum state. Using Routh's method of ignorable coordinates it is shown that the quantum potential energy of particle interaction that represents quantum effects in this model may be regarded as the kinetic energy of additional 'concealed' freedoms. The method brings an alternative perspective to Planck's constant, which plays the role of a hidden variable, and to the canonical quantization procedure, since what is termed 'kinetic energy' in quantum mechanics may be regarded literally as energy due to motion.

## 1 Potential energy as kinetic energy

In the late 19[th] century some natural philosophers espoused the view that potential energy should be understood as a phenomenological quantity whose form could be divined from observation but whose nature resided ultimately in motion, which should more appropriately be represented by kinetic energy. According to J.J. Thompson, who wrote a book from this perspective [1],

> *'This view which regards all potential energy as really kinetic has the advantage of keeping before us the idea that it is one of the objects of Physical Science to explain natural phenomena by means of the properties of matter in motion. When we have done this we have got a complete physical explanation of any phenomenon…It is not so however when we explain the phenomenon as due to changes in the potential energy of the system; for potential energy cannot be said, in the strict sense of the term, to explain anything. It does little more than embody the results of experiments in a form suitable for mathematical investigations.'*

Proceeding from the assumption that physical phenomena should be regarded as occurring within the framework of space and time, one of the problems perceived with potential energy was the difficulty of localizing it, that is, of apportioning it to specific regions of space. It was proposed to treat potential energy as the representative in the domain of 'visible' coordinates of the kinetic energy of 'concealed' motions whose character cannot be fully specified due, for example, to their complexity or to incomplete information. As observed by Webster [2], 'a system devoid of potential energy would seem to possess it, if we were in ignorance of the existence of the…motions in it.' The matter whose motion constitutes the kinetic energy may be either part of the system or a system external to it (for a 19[th] century physicist the latter might be the ether).

An implementation of this programme grounded in analytical dynamics was provided by the method of ignorable coordinates due to Routh [3]. This method relates to systems for which a subset of the coordinates do not appear in the Lagrangian while the corresponding velocities do. In the context of representing potential energy by kinetic energy Routh's technique featured in several texts of the period (e.g., [2,4]) but this physical application tended to disappear from later treatments, which emphasized just its mathematical merits (e.g., [5-8]; see [9] for hydrodynamic applications).

Consider a mechanical system comprising two components, one for which the coordinates $q_i(t)$ are observable or 'visible', and the other for which the coordinates $Q_i(t)$ are 'concealed', their presence being manifested indirectly through the visible coordinates (we take $i,j,… = 1,2,3$ for each component but the extension to arbitrary dimensions for either is straightforward). The Lagrangian is assumed to be a homogeneous quadratic function of the velocities determined by all the freedoms: $L = T_q + T_Q$ or

$$L(q,\dot{q},\dot{Q}) = \tfrac{1}{2} B_{ij}(q)\dot{q}_i \dot{q}_j + \tfrac{1}{2} A_{ij}(q)\dot{Q}_i \dot{Q}_j, \tag{1.1}$$

where the symmetric matrices $A_{ij}, B_{ij}$ depend only on $q_i$ and are invertible. It may appear unusual that a function of $q_i$ appears as a coefficient in the second term in (1.1) but this stems from the standard form for the kinetic energy; denoting the configuration space coordinates collectively by $\xi_\mu = (q_i, Q_i)$, $\mu = 1,...,6$, we have $L = \tfrac{1}{2} m g_{\mu\nu}(\xi) \dot{\xi}_\mu \dot{\xi}_\nu$. The aim is to show that the dynamics of the system may be derived equally from a modified Lagrangian in which the kinetic term $T_Q$ is assimilable to potential energy.

The canonical momentum corresponding to $Q_i$ is

$$P_i = \frac{\partial L}{\partial \dot{Q}_i} = A_{ij}(q)\dot{Q}_j. \tag{1.2}$$

The Euler-Lagrange equations for $Q_i$ imply that $P_i$ is constant. We may then eliminate $\dot{Q}_i$ from $L$ by solving (1.2) to give $\dot{Q}_i = A^{-1}{}_{ij} P_j$ where, evaluating (1.2) at $t = 0$, $P_i$ is determined by the initial conditions via the formula $P_i = A_{ij}(q_0)\dot{Q}_{0j}$. Now, define the modified Lagrangian function (the 'Routhian'):

$$L'(q,\dot{q},P) = L - P_i \dot{Q}_i = \tfrac{1}{2} B_{ij}(q)\dot{q}_i \dot{q}_j - \tfrac{1}{2} A^{-1}{}_{ij}(q) P_i P_j. \tag{1.3}$$

Using (1.2), we may show that the Euler-Lagrange equations for $q_i$ derived from $L'$,

$$\frac{d}{dt}\left(\frac{\partial L'}{\partial \dot{q}_i}\right) = \frac{\partial L'}{\partial q_i}, \tag{1.4}$$

coincide with those derived from $L$ but now, as promised, the motion proceeds as if this component of the system possesses potential energy $V_q = T_Q$: $L' = T_q - V_q$. The kinetic energy of the $Q$-system may therefore be regarded as potential energy of the $q$-system.

As regards the concealed motion $Q_i(t)$, the (negative of the) function $L'$ plays the role of the Hamiltonian; Hamilton's equations give

$$\dot{Q}_i = -\frac{\partial L'}{\partial P_i} = A^{-1}{}_{ij}(q)P_j, \quad \dot{P}_i = \frac{\partial L'}{\partial Q_i} = 0. \tag{1.5}$$

The motion of this component of the total system is therefore determined once $q_i(t)$ has been found. Information on the concealed motion is contained in the emergent potential energy via the constants $P_i$, which are fixed by the initial concealed velocities. It is noted that we could develop the entire theory, including the visible motion, in Hamiltonian terms but we adhere to the historical analytic route, which is advantageous in connection with certain applications such as Noether's theorem.

It is clear that any potential energy may be treated as the visible effect of concealed motions, to which the forces between visible segments of matter may therefore be attributed. However, it should be observed that, while Routh's demonstration offers an alternative perspective on potential energy as shorthand for hidden motion, it entails a simple reformulation of variational theory and a correspondingly modest new insight. For example, as argued by Whittaker [10], it does not appear to assist in solving the problem of localizing potential energy. Despite this shortcoming, there is an area of physical science where it is of interest to seek to apply these ideas. It is known that in quantum mechanics a system may be modelled in terms of a process where the kinetic energy operator is associated partly with kinetic energy and partly with potential energy. We shall show that quantum evolution may be viewed as the outcome of visible and concealed motions, without recourse to potential energy.

**2 Application to quantum mechanics**

Routh's method may be extended straightforwardly to a continuum if the coordinates identifying the particle freedoms are suitably ordered, such as occurs in the limiting case of a lattice. The theory may then be applied to a fluid in its Lagrangian formulation. This provides a natural forum for the application of Routh's scheme to quantum theory for, as is well known, the latter admits a description as a kind of hydrodynamics (see [11] and references therein). Writing the wavefunction as $\psi(x,t) = |\psi|\exp(iS/\hbar)$, the density and velocity fields in the Eulerian description of the quantum fluid are given by $\rho(x,t) = |\psi|^2$ and $v_i(x,t) = m^{-1}\partial S/\partial x_i$, respectively. In the Lagrangian picture of a fluid, the freedoms of an element of a continuous medium can be specified by the coordinates $a_i$ of a point occupied by the element at a fixed time (taken as $t = 0$) and the coordinates $q_i(a,t)$ of its position as it moves (with $q_i(a,0) = a_i$). It has been shown by the author that quantal evolution may be computed independently of the wavefunction by representing the quantum state as the function $q_i$ in such a model (where the initial conditions are correlated with the initial wavefunction) [11-16]. This quantum Lagrangian picture, which provides a formula for the time-dependent wavefunction in terms of the trajectories, is the version of quantum hydrodynamics to which we shall apply Routh's method.

Hitherto, a dualist approach to quantum hydrodynamics has been taken in which the quantum effects are attributed to the potential energy of interaction of neighbouring particles in the fluid. Representing quantum states as spatial displacements allows us to show how the interaction energy may be grounded in the kinetic energy of concealed motion, following Routh's procedure. To this end, we represent a quantum system using a two-fluid model in which space is filled with two interpenetrating fluids, one having 'visible' coordinates $q_i(a,t)$ and the other having 'concealed' coordinates $Q_i(a,t)$ (where in general $Q_i(a,0) \neq a_i$). Each space point supports a fluid particle of each species (for details of the relevant aspects of multiphase fluid theory see [17,18]).

For the visible fluid, let $\rho_0(a)$ be the initial number density (this is normalized, $\int \rho_0(a) d^3a = 1$, and will be identified with the initial quantal probability density). Then, introducing a mass parameter $m$, the mass of an elementary volume $d^3a$ attached to the point with coordinates $a_i$ is given by $m\rho_0(a)d^3a$. The conservation of the mass of the element in the course of its motion is expressed through the relation $m\rho(q(a,t),t)d^3q(a,t) = m\rho_0(a)d^3a$, which determines the density at time $t$: $\rho(q,t) = J^{-1}(a,t)\rho_0(a)$ where $J$ is the Jacobian of the transformation between the two sets of coordinates. The reference density of the concealed fluid is denoted $P_0(a)$. This fluid may also be regarded as locally conserved but we shall not make use of this aspect.

Consider a class of theories where the Lagrangian is the following combination of the kinetic energies of the two fluids, an obvious generalization of the discrete particle Lagrangian (1.1):

$$L[q,\dot{q},\dot{Q}] = \int \tfrac{1}{2} m[\rho_0(a)B_{ij}(a,q)\dot{q}_i(a)\dot{q}_j(a) + P_0(a)A_{ij}(a,q)\dot{Q}_i(a)\dot{Q}_j(a)]d^3a \quad (2.1)$$

where $\dot{q}_i(a) = \partial q_i(a,t)/\partial t$ and $\dot{Q}_i(a) = \partial Q_i(a,t)/\partial t$. This model exhibits possibilities not available in the discrete case. For example, the matrices $A_{ij}(a,q), B_{ij}(a,q)$ may depend on the deformation coefficients $\partial q_i/\partial a_j$. There are several ways of achieving our end depending on how the various functions are identified and we make the following choices: $P_0(a) = \rho_0(a)$, $A_{ij} = u_{0k}(a)u_{0k}(a)(u_l(q)u_l(q))^{-1}\delta_{ij}$, $B_{ij} = \delta_{ij}$. Here $u_{0i}(a) = \partial \log \rho_0 / \partial a_i$ and $u_i(q) = \partial \log \rho / \partial q_i$ with $\partial/\partial q_i = J^{-1}J_{ij}\partial/\partial a_j$, $J_{ij}$ being the cofactor of $\partial q_i/\partial a_j$. Then

$$L[q,\dot{q},\dot{Q}] = \int \tfrac{1}{2} m\rho_0(a)[\dot{q}_i(a)\dot{q}_i(a) + u_{0k}u_{0k}(u_j u_j)^{-1}\dot{Q}_i(a)\dot{Q}_i(a)]d^3a. \quad (2.2)$$

The canonical field momenta for the concealed coordinates are

$$P_i(a) = \frac{\delta L}{\delta \dot{Q}_i} = m\rho_0(a)u_{0k}u_{0k}(u_i u_i)^{-1}\dot{Q}_i(a). \quad (2.3)$$

By the Euler-Lagrange equations these are constants of the motion and equal the initial momentum density: $P_i(a) = m\rho_0(a)\dot{Q}_{0i}(a)$. The modified Lagrangian is

$$L'[q,\dot{q},P] = L - \int P_i(a)\dot{Q}_i(a)d^3a$$
$$= \int [\tfrac{1}{2}m\rho_0(a)\dot{q}_i(a)\dot{q}_i(a) \qquad (2.4)$$
$$- (2m\rho_0(a)u_{0k}(a)u_{0k}(a))^{-1}u_i(q)u_i(q)P_i(a)P_i(a)]d^3a.$$

To complete the story, we make the following choice for the magnitude of the initial velocity of the concealed flow, one that characterizes quantum mechanical evolution:

$$\dot{Q}_{0i}(a)\dot{Q}_{0i}(a) = \hbar^2 u_{0k}(a)u_{0k}(a)/4m^2. \qquad (2.5)$$

For then

$$L'[q,\dot{q}] = \int \rho_0(a)[\tfrac{1}{2}m\dot{q}_i(a)\dot{q}_i(a) - (\hbar^2/8m)u_i(q)u_i(q)]d^3a \qquad (2.6)$$

and the effects due to the concealed motion are now attributed to the quantum potential energy $(\hbar^2/8m)\rho_0 u_i u_i$ introduced previously [11]. The Euler-Lagrange equations for the visible motion $q_i(a,t)$ derived from $L'$ imply a nonlinear equation that, under the assumption of initial potential flow (which is preserved by the subsequent flow), translates into the free Schrödinger equation $i\hbar\partial\psi/\partial t = -\hbar^2\nabla^2\psi/2m$ in the Eulerian picture. Here $\psi(x,t) = \sqrt{\rho}\exp(iS/\hbar)$ where $\rho(x,t) = J^{-1}(a,t)\rho_0(a)\big|_{a(x=q,t)}$ and $v_i(x,t) = m^{-1}\partial S/\partial x_i = \dot{q}_i(a,t)\big|_{a(x=q,t)}$.

As expected, the function $-L'[q,\dot{q},P]$ provides the Hamiltonian for the concealed motion. Hamilton's equations are

$$\dot{Q}_i(a) = -\frac{\delta L'}{\delta P_i(a)} = (u_{0k}u_{0k})^{-1}u_j u_j \dot{Q}_{0i}(a), \quad \dot{P}_i(a) = \frac{\delta L'}{\delta Q_i(a)} = 0, \qquad (2.7)$$

which have the solution

$$Q_i(a,t) = Q_{0i}(a) + (u_{0k}(a)u_{0k}(a))^{-1}\dot{Q}_{0i}(a)\int_0^t u_j(q(a,t))u_j(q(a,t))dt. \qquad (2.8)$$

In epochs for which $\dot{q}_i\dot{q}_i \gg \dot{Q}_{0i}\dot{Q}_i$, or $|\dot{q}_i| \gg (\hbar/2m)|u_i(q)|$, the quantum component of the energy is negligible and the motion becomes classical.

The equation of motion (2.7) for the concealed freedoms is a conservation equation for the current $(u_j u_j)^{-1}\dot{Q}_i(a)$. Defining the velocity field $V_i(x,t) = J^{-1}(a,t)\dot{Q}_i(a,t)\big|_{a(x=q,t)}$ and using the standard conversion formulas between the Lagrangian and Eulerian descriptions [14], the equivalent Eulerian continuity equation is

$$\frac{\partial}{\partial t}\left(\frac{V_i}{u_j u_j}\right) + \frac{\partial}{\partial x_k}\left(\frac{V_i v_k}{u_j u_j}\right) = 0 \qquad (2.9)$$

where $u_i(x) = \partial \log \rho(x)/\partial x_i$. This equation is appended to the Schrödinger equation and may be solved for $V_i$ in place of (2.7) once the wavefunction, and hence $v_i(x)$ and $u_i(x)$, has been determined by solving either the Euler-Lagrange equations for $q_i(a,t)$ or the Schrödinger equation directly.

The Lagrangian (2.2), comprising just kinetic energy, gives the total energy of the system. It is easy to derive the equivalent form in the Eulerian variables:

$$\begin{aligned} H &= \int \tfrac{1}{2} m \rho_0(a)[\dot{q}_i(a)\dot{q}_i(a) + u_{0k}u_{0k}(u_j u_j)^{-1} \dot{Q}_i(a)\dot{Q}_i(a)] d^3a \\ &= \int \left( \frac{1}{2m} \rho \frac{\partial S}{\partial x_i}\frac{\partial S}{\partial x_i} + \frac{\hbar^2}{8m\rho}\frac{\partial \rho}{\partial x_i}\frac{\partial \rho}{\partial x_i} \right) d^3x \\ &= \int \psi^* \left( -\frac{\hbar^2}{2m} \right) \frac{\partial^2}{\partial x_i \partial x_i} \psi \, d^3 x. \end{aligned} \qquad (2.10)$$

In the last step we have performed an integration by parts to get the second-order operator. We thus obtain the quantum mechanical formula for the mean kinetic energy; what is termed 'kinetic energy' in quantum mechanics may be regarded literally as energy due to motion. It may be expressed in the standard form

$$H = \int \tfrac{1}{2} m \rho_0(a) g_{\mu\nu}(a,\xi) \dot{\xi}_\mu(a) \dot{\xi}_\nu(a) \, d^3 a. \qquad (2.11)$$

The above analysis may be broadened to include an external potential $V$ by replacing $(\hbar^2/8m)u_i u_i$ by $(\hbar^2/8m)u_i u_i + V$ or by introducing a separate hidden kinetic energy to represent $V$. A corresponding extension can be made to incorporate an external vector potential. Generalizing the configuration space to an $n$-dimensional Riemannian manifold we may extend the treatment to embrace a wide variety of systems where the quantum state is represented by Lagrangian coordinates, including many bodies, spin ½ and fields [11-16]. Note that in the latter cases the arena for the flow is the configuration space of the system and not physical space. The Riemannian extension also allows us to apply the approach to other field theories such as electromagnetism where a quantum potential energy-like term appears in the Lagrangian in a fluid-dynamical representation of Maxwell's equations [12].

### 3 Comments

We have shown that quantum mechanical evolution, treated as the unravelling of a time-dependent coordinate transformation $a_i \to q_i(a,t)$, may be regarded as originating in motions whose full details are hidden and may in part be attributed to potential energy. The designation of the paths $q_i(a,t)$ as 'visible' is justified in the quantum context since these are the variables observed in position measurements, which build the statistical distribution $|\psi(x,t)|^2 = \rho(x = q, t)$. Similarly, the variables $Q_i(a,t)$ deserve the epithet 'concealed' as they are manifested only indirectly through the visible coordinates, which however provide no access to them. In particular, Planck's constant, which enters the dynamical theory through the concealed variables (cf. (2.5)) and thus contributes to defining the state of the whole system, may be regarded as a 'hidden variable'.

A corollary is that we may establish by this means consistency in the use of language; when the phrase 'kinetic energy' is employed in quantum mechanics, in connection with the operator obtained by the canonical quantization procedure, we may refer to a model of a quantum system for which the terminology is justified according to our intuitive understanding of kinetic energy as signifying and quantifying matter in motion.

This analysis also provides a kinetic basis for the quantum potential of de Broglie and Bohm ($=-\hbar^2\nabla^2\sqrt{\rho}/2m\sqrt{\rho}$), which is the representative of the quantum potential energy in the de Broglie-Bohm particle law of motion. Note that the representation of the wavefunction as a continuum of fluid particles is independent of the assumption of the de Broglie-Bohm model that one of the lines of flow is populated by a corpuscle, a notion that is supported by but is not implicit in the fluid model. One could of course apply Routh's method directly to the particle Lagrangian containing the quantum potential. This could provide justification for Einstein's attempted hidden-variable theory – in which the quantum potential is treated as kinetic energy – although it has been shown that this theory is generally untenable [19].

The model we have presented has the status of a 'proof of concept'. In this primitive formulation it provides no explanation for the form of the density-dependent factor in the concealed kinetic energy or for the value of Planck's constant. There are considerable opportunities for modifying the model. We could, for example, choose $P_0(a)$ differently if a corresponding alteration is made in the definition of the initial concealed velocity. And instead of interpreting the concealed variables as pertaining to a separate physical system, we might let them represent internal freedoms (such as rotation) or spatial coordinates in higher dimensions. In the latter case the theory is closely connected with the Kaluza-Klein programme. An interesting aspect of the concealed motion, however it is conceived, is that it is described by a continuity equation (2.9) corresponding to the quantum flow $v_i$ but where the density differs from the quantum (probability) expression.

[1] J.J. Thompson, *Applications of Dynamics to Physics and Chemistry* (Macmillan, London, 1888) p. 15.
[2] A.G. Webster, *The Dynamics of Particles and of Rigid, Elastic, and Fluid Bodies* (Teubner, Leipzig, 1904) sec. 48.
[3] E.J. Routh, *A Treatise on the Stability of Motion* (Macmillan, London, 1877) p. 60.
[4] H.M. Macdonald, *Electric Waves* (Cambridge University Press, Cambridge, 1902) chap. 5.
[5] E.T. Whittaker, *A Treatise on the Analytical Dynamics of Particles and Rigid Bodies*, 4th edition (Cambridge University Press, Cambridge, 1952) p. 54.
[6] W.E. Byerly, *An Introduction to the Use of Generalized Coordinates in Mechanics and Physics* (Ginn and Co., Boston, 1944) chap. 2.
[7] H. Goldstein, *Classical Mechanics* (Addison-Wesley, Reading, 1950) chap. 7.
[8] L.A. Pars, *A Treatise on Analytical Dynamics* (Heinemann, London, 1965) chap. 10.

[9] H. Lamb, *Hydrodynamics*, 6th edition (Cambridge University Press, Cambridge, 1932) chap. 6.
[10] E.T. Whittaker, *From Euclid to Eddington* (Cambridge University Press, Cambridge, 1949) p. 84.
[11] P. Holland, *Ann. Phys. (NY)* 315, 503 (2005).
[12] P. Holland, *Proc. R. Soc. A* 461, 3659 (2005).
[13] P. Holland, in *Quantum Trajectories*, ed. P. Chattaraj (Taylor & Francis/CRC, Boca Raton, 2010) chap. 5.
[14] P. Holland, in *Concepts and Methods in Modern Theoretical Chemistry: Statistical Mechanics*, eds. S.K. Ghosh and P.K. Chattaraj (Taylor & Francis/CRC, Boca Raton, 2013) chap. 4.
[15] P. Holland, *Found. Phys.* 36, 1 (2006).
[16] P. Holland, *Int. J. Theor. Phys.* 51, 667 (2012).
[17] J.A. Guerst, *Physica A* 135, 455 (1986).
[18] P. Holland, *J. Phys. A: Math. Theor.* 42, 075307 (2009).
[19] P. Holland, *Found. Phys.* 35, 177 (2005).